\documentstyle[12pt,aps,epsfig]{revtex}
\renewcommand {\baselinestretch}{1.}
\def\PsfigVersion{1.9}
\ifx\undefined\psfig\else \fi

\let\LaTeXAtSign=\@
\let\@=\relax
\edef\psfigRestoreAt{\catcode`\@=\number\catcode`@\relax}
\catcode`\@=11\relax
\newwrite\@unused
\def\ps@typeout#1{{\let\protect\string\immediate\write\@unused{#1}}}
\ps@typeout{psfig/tex \PsfigVersion}

\def\figurepath{./}

\def\@nnil{\@nil}
\def\@empty{}
\def\@psdonoop#1\@@#2#3{}
\def\@psdo#1:=#2\do#3{\edef\@psdotmp{#2}\ifx\@psdotmp\@empty \else
    \expandafter\@psdoloop#2,\@nil,\@nil\@@#1{#3}\fi}
\def\@psdoloop#1,#2,#3\@@#4#5{\def#4{#1}\ifx #4\@nnil \else
       #5\def#4{#2}\ifx #4\@nnil \else#5\@ipsdoloop #3\@@#4{#5}\fi\fi}
\def\@ipsdoloop#1,#2\@@#3#4{\def#3{#1}\ifx #3\@nnil 
       \let\@nextwhile=\@psdonoop \else
      #4\relax\let\@nextwhile=\@ipsdoloop\fi\@nextwhile#2\@@#3{#4}}
\def\@tpsdo#1:=#2\do#3{\xdef\@psdotmp{#2}\ifx\@psdotmp\@empty \else
    \@tpsdoloop#2\@nil\@nil\@@#1{#3}\fi}
\def\@tpsdoloop#1#2\@@#3#4{\def#3{#1}\ifx #3\@nnil 
       \let\@nextwhile=\@psdonoop \else
      #4\relax\let\@nextwhile=\@tpsdoloop\fi\@nextwhile#2\@@#3{#4}}
\ifx\undefined\fbox
\newdimen\fboxrule
\newdimen\fboxsep
\newdimen\ps@tempdima
\newbox\ps@tempboxa
\long\def\fbox#1{\leavevmode\setbox\ps@tempboxa\hbox{#1}\ps@tempdima\fboxrule
    \advance\ps@tempdima \fboxsep \advance\ps@tempdima \dp\ps@tempboxa
   \hbox{\lower \ps@tempdima\hbox
  {\vbox{\hrule height \fboxrule
          \hbox{\vrule width \fboxrule \hskip\fboxsep
          \vbox{\vskip\fboxsep \box\ps@tempboxa\vskip\fboxsep}\hskip 
                 \fboxsep\vrule width \fboxrule}
                 \hrule height \fboxrule}}}}
\fi
\newread\ps@stream
\newif\ifnot@eof       % continue looking for the bounding box?
\newif\if@noisy        % report what you're making?
\newif\if@atend        % %%BoundingBox: has (at end) specification
\newif\if@psfile       % does this look like a PostScript file?
{\catcode`\%=12\global\gdef\epsf@start{%!}}
\def\epsf@PS{PS}
\def\epsf@getbb#1{%
\openin\ps@stream=#1
\ifeof\ps@stream\ps@typeout{Error, File #1 not found}\else
   {\not@eoftrue \chardef\other=12
    \def\do##1{\catcode`##1=\other}\dospecials \catcode`\ =10
    \loop
       \if@psfile
	  \read\ps@stream to \epsf@fileline
       \else{
	  \obeyspaces
          \read\ps@stream to \epsf@tmp\global\let\epsf@fileline\epsf@tmp}
       \fi
       \ifeof\ps@stream\not@eoffalse\else
       \if@psfile\else
       \expandafter\epsf@test\epsf@fileline:. \\%
       \fi
          \expandafter\epsf@aux\epsf@fileline:. \\%
       \fi
   \ifnot@eof\repeat
   }\closein\ps@stream\fi}%
\long\def\epsf@test#1#2#3:#4\\{\def\epsf@testit{#1#2}
			\ifx\epsf@testit\epsf@start\else
\ps@typeout{Warning! File does not start with `\epsf@start'.  It may not be a PostScript file.}
			\fi
			\@psfiletrue} % don't test after 1st line
{\catcode`\%=12\global\let\epsf@percent=%\global\def\epsf@bblit{%BoundingBox}}
\long\def\epsf@aux#1#2:#3\\{\ifx#1\epsf@percent
   \def\epsf@testit{#2}\ifx\epsf@testit\epsf@bblit
	\@atendfalse
        \epsf@atend #3 . \\%
	\if@atend	
	   \if@verbose{
		\ps@typeout{psfig: found `(atend)'; continuing search}
	   }\fi
        \else
        \epsf@grab #3 . . . \\%
        \not@eoffalse
        \global\no@bbfalse
        \fi
   \fi\fi}%
\def\epsf@grab #1 #2 #3 #4 #5\\{%
   \global\def\epsf@llx{#1}\ifx\epsf@llx\empty
      \epsf@grab #2 #3 #4 #5 .\\\else
   \global\def\epsf@lly{#2}%
   \global\def\epsf@urx{#3}\global\def\epsf@ury{#4}\fi}%
\def\epsf@atendlit{(atend)} 
\def\epsf@atend #1 #2 #3\\{%
   \def\epsf@tmp{#1}\ifx\epsf@tmp\empty
      \epsf@atend #2 #3 .\\\else
   \ifx\epsf@tmp\epsf@atendlit\@atendtrue\fi\fi}

\chardef\psletter = 11 % won't conflict with \begin{letter} now...
\chardef\other = 12

\newif \ifdebug %%% turn me on to see TeX hard at work ...
\newif\ifc@mpute %%% don't need to compute some values
\c@mputetrue % but assume that we do

\let\then = \relax
\def\r@dian{pt }
\let\r@dians = \r@dian
\let\dimensionless@nit = \r@dian
\let\dimensionless@nits = \dimensionless@nit
\def\internal@nit{sp }
\let\internal@nits = \internal@nit
\newif\ifstillc@nverging
\def \Mess@ge #1{\ifdebug \then \message {#1} \fi}

{ %%% Things that need abnormal catcodes %%%
	\catcode `\@ = \psletter
	\gdef \nodimen {\expandafter \n@dimen \the \dimen}
	\gdef \term #1 #2 #3%
	       {\edef \t@ {\the #1}%%% freeze parameter 1 (count, by value)
		\edef \t@@ {\expandafter \n@dimen \the #2\r@dian}%
		\t@rm {\t@} {\t@@} {#3}%
	       }
	\gdef \t@rm #1 #2 #3%
	       {{%
		\count 0 = 0
		\dimen 0 = 1 \dimensionless@nit
		\dimen 2 = #2\relax
		\Mess@ge {Calculating term #1 of \nodimen 2}%
		\loop
		\ifnum	\count 0 < #1
		\then	\advance \count 0 by 1
			\Mess@ge {Iteration \the \count 0 \space}%
			\Multiply \dimen 0 by {\dimen 2}%
			\Mess@ge {After multiplication, term = \nodimen 0}%
			\Divide \dimen 0 by {\count 0}%
			\Mess@ge {After division, term = \nodimen 0}%
		\repeat
		\Mess@ge {Final value for term #1 of 
				\nodimen 2 \space is \nodimen 0}%
		\xdef \Term {#3 = \nodimen 0 \r@dians}%
		\aftergroup \Term
	       }}
	\catcode `\p = \other
	\catcode `\t = \other
	\gdef \n@dimen #1pt{#1} %%% throw away the ``pt''
}

\def \Divide #1by #2{\divide #1 by #2} %%% just a synonym

\def \Multiply #1by #2%%% allows division of a dimen by a dimen
       {{%%% should really freeze parameter 2 (dimen, passed by value)
	\count 0 = #1\relax
	\count 2 = #2\relax
	\count 4 = 65536
	\Mess@ge {Before scaling, count 0 = \the \count 0 \space and
			count 2 = \the \count 2}%
	\ifnum	\count 0 > 32767 %%% do our best to avoid overflow
	\then	\divide \count 0 by 4
		\divide \count 4 by 4
	\else	\ifnum	\count 0 < -32767
		\then	\divide \count 0 by 4
			\divide \count 4 by 4
		\else
		\fi
	\fi
	\ifnum	\count 2 > 32767 %%% while retaining reasonable accuracy
	\then	\divide \count 2 by 4
		\divide \count 4 by 4
	\else	\ifnum	\count 2 < -32767
		\then	\divide \count 2 by 4
			\divide \count 4 by 4
		\else
		\fi
	\fi
	\multiply \count 0 by \count 2
	\divide \count 0 by \count 4
	\xdef \product {#1 = \the \count 0 \internal@nits}%
	\aftergroup \product
       }}

\def\r@duce{\ifdim\dimen0 > 90\r@dian \then   % sin(x+90) = sin(180-x)
		\multiply\dimen0 by -1
		\advance\dimen0 by 180\r@dian
		\r@duce
	    \else \ifdim\dimen0 < -90\r@dian \then  % sin(-x) = sin(360+x)
		\advance\dimen0 by 360\r@dian
		\r@duce
		\fi
	    \fi}

\def\Sine#1%
       {{%
	\dimen 0 = #1 \r@dian
	\r@duce
	\ifdim\dimen0 = -90\r@dian \then
	   \dimen4 = -1\r@dian
	   \c@mputefalse
	\fi
	\ifdim\dimen0 = 90\r@dian \then
	   \dimen4 = 1\r@dian
	   \c@mputefalse
	\fi
	\ifdim\dimen0 = 0\r@dian \then
	   \dimen4 = 0\r@dian
	   \c@mputefalse
	\fi
	\ifc@mpute \then
		\divide\dimen0 by 180
		\dimen0=3.141592654\dimen0
		\dimen 2 = 3.1415926535897963\r@dian %%% a well-known constant
		\divide\dimen 2 by 2 %%% we only deal with -pi/2 : pi/2
		\Mess@ge {Sin: calculating Sin of \nodimen 0}%
		\count 0 = 1 %%% see power-series expansion for sine
		\dimen 2 = 1 \r@dian %%% ditto
		\dimen 4 = 0 \r@dian %%% ditto
		\loop
			\ifnum	\dimen 2 = 0 %%% then we've done
			\then	\stillc@nvergingfalse 
			\else	\stillc@nvergingtrue
			\fi
			\ifstillc@nverging %%% then calculate next term
			\then	\term {\count 0} {\dimen 0} {\dimen 2}%
				\advance \count 0 by 2
				\count 2 = \count 0
				\divide \count 2 by 2
				\ifodd	\count 2 %%% signs alternate
				\then	\advance \dimen 4 by \dimen 2
				\else	\advance \dimen 4 by -\dimen 2
				\fi
		\repeat
	\fi		
			\xdef \sine {\nodimen 4}%
       }}

\def\Cosine#1{\ifx\sine\UnDefined\edef\Savesine{\relax}\else
		             \edef\Savesine{\sine}\fi
	{\dimen0=#1\r@dian\advance\dimen0 by 90\r@dian
	 \Sine{\nodimen 0}
	 \xdef\cosine{\sine}
	 \xdef\sine{\Savesine}}}	      
\def\psdraft{
	\def\@psdraft{0}
}
\def\psfull{
	\def\@psdraft{100}
}

\psfull

\newif\if@scalefirst
\def\psscalefirst{\@scalefirsttrue}
\def\psrotatefirst{\@scalefirstfalse}
\psrotatefirst

\newif\if@draftbox
\def\psnodraftbox{
	\@draftboxfalse
}
\def\psdraftbox{
	\@draftboxtrue
}
\@draftboxtrue

\newif\if@prologfile
\newif\if@postlogfile
\def\pssilent{
	\@noisyfalse
}
\def\psnoisy{
	\@noisytrue
}
\psnoisy
\newif\if@bbllx
\newif\if@bblly
\newif\if@bburx
\newif\if@bbury
\newif\if@height
\newif\if@width
\newif\if@rheight
\newif\if@rwidth
\newif\if@angle
\newif\if@clip
\newif\if@verbose
\def\@p@@sclip#1{\@cliptrue}

\newif\if@decmpr

\def\@p@@sfigure#1{\def\@p@sfile{null}\def\@p@sbbfile{null}
	        \openin1=#1.bb
		\ifeof1\closein1
	        	\openin1=\figurepath#1.bb
			\ifeof1\closein1
			        \openin1=#1
				\ifeof1\closein1%
				       \openin1=\figurepath#1
					\ifeof1
					   \ps@typeout{Error, File #1 not found}
						\if@bbllx\if@bblly
				   		\if@bburx\if@bbury
			      				\def\@p@sfile{#1}%
			      				\def\@p@sbbfile{#1}%
							\@decmprfalse
				  	   	\fi\fi\fi\fi
					\else\closein1
				    		\def\@p@sfile{\figurepath#1}%
				    		\def\@p@sbbfile{\figurepath#1}%
						\@decmprfalse
	                       		\fi%
			 	\else\closein1%
					\def\@p@sfile{#1}
					\def\@p@sbbfile{#1}
					\@decmprfalse
			 	\fi
			\else
				\def\@p@sfile{\figurepath#1}
				\def\@p@sbbfile{\figurepath#1.bb}
				\@decmprtrue
			\fi
		\else
			\def\@p@sfile{#1}
			\def\@p@sbbfile{#1.bb}
			\@decmprtrue
		\fi}

\def\@p@@sfile#1{\@p@@sfigure{#1}}

\def\@p@@sbbllx#1{
		\@bbllxtrue
		\dimen100=#1
		\edef\@p@sbbllx{\number\dimen100}
}
\def\@p@@sbblly#1{
		\@bbllytrue
		\dimen100=#1
		\edef\@p@sbblly{\number\dimen100}
}
\def\@p@@sbburx#1{
		\@bburxtrue
		\dimen100=#1
		\edef\@p@sbburx{\number\dimen100}
}
\def\@p@@sbbury#1{
		\@bburytrue
		\dimen100=#1
		\edef\@p@sbbury{\number\dimen100}
}
\def\@p@@sheight#1{
		\@heighttrue
		\dimen100=#1
   		\edef\@p@sheight{\number\dimen100}
}
\def\@p@@swidth#1{
		\@widthtrue
		\dimen100=#1
		\edef\@p@swidth{\number\dimen100}
}
\def\@p@@srheight#1{
		\@rheighttrue
		\dimen100=#1
		\edef\@p@srheight{\number\dimen100}
}
\def\@p@@srwidth#1{
		\@rwidthtrue
		\dimen100=#1
		\edef\@p@srwidth{\number\dimen100}
}
\def\@p@@sangle#1{
		\@angletrue
		\edef\@p@sangle{#1} %\number\dimen100}
}
\def\@p@@ssilent#1{ 
		\@verbosefalse
}
\def\@p@@sprolog#1{\@prologfiletrue\def\@prologfileval{#1}}
\def\@p@@spostlog#1{\@postlogfiletrue\def\@postlogfileval{#1}}
\def\@cs@name#1{\csname #1\endcsname}
\def\@setparms#1=#2,{\@cs@name{@p@@s#1}{#2}}
\def\ps@init@parms{
		\@bbllxfalse \@bbllyfalse
		\@bburxfalse \@bburyfalse
		\@heightfalse \@widthfalse
		\@rheightfalse \@rwidthfalse
		\def\@p@sbbllx{}\def\@p@sbblly{}
		\def\@p@sbburx{}\def\@p@sbbury{}
		\def\@p@sheight{}\def\@p@swidth{}
		\def\@p@srheight{}\def\@p@srwidth{}
		\def\@p@sangle{0}
		\def\@p@sfile{} \def\@p@sbbfile{}
		\def\@p@scost{10}
		\def\@sc{}
		\@prologfilefalse
		\@postlogfilefalse
		\@clipfalse
		\if@noisy
			\@verbosetrue
		\else
			\@verbosefalse
		\fi
}
\def\parse@ps@parms#1{
	 	\@psdo\@psfiga:=#1\do
		   {\expandafter\@setparms\@psfiga,}}
\newif\ifno@bb
\def\bb@missing{
	\if@verbose{
		\ps@typeout{psfig: searching \@p@sbbfile \space  for bounding box}
	}\fi
	\no@bbtrue
	\epsf@getbb{\@p@sbbfile}
        \ifno@bb \else \bb@cull\epsf@llx\epsf@lly\epsf@urx\epsf@ury\fi
}	
\def\bb@cull#1#2#3#4{
	\dimen100=#1 bp\edef\@p@sbbllx{\number\dimen100}
	\dimen100=#2 bp\edef\@p@sbblly{\number\dimen100}
	\dimen100=#3 bp\edef\@p@sbburx{\number\dimen100}
	\dimen100=#4 bp\edef\@p@sbbury{\number\dimen100}
	\no@bbfalse
}
\newdimen\p@intvaluex
\newdimen\p@intvaluey
\def\rotate@#1#2{{\dimen0=#1 sp\dimen1=#2 sp
		  \global\p@intvaluex=\cosine\dimen0
		  \dimen3=\sine\dimen1
		  \global\advance\p@intvaluex by -\dimen3
		  \global\p@intvaluey=\sine\dimen0
		  \dimen3=\cosine\dimen1
		  \global\advance\p@intvaluey by \dimen3
		  }}
\def\compute@bb{
		\no@bbfalse
		\if@bbllx \else \no@bbtrue \fi
		\if@bblly \else \no@bbtrue \fi
		\if@bburx \else \no@bbtrue \fi
		\if@bbury \else \no@bbtrue \fi
		\ifno@bb \bb@missing \fi
		\ifno@bb \ps@typeout{FATAL ERROR: no bb supplied or found}
			\no-bb-error
		\fi
		\count203=\@p@sbburx
		\count204=\@p@sbbury
		\advance\count203 by -\@p@sbbllx
		\advance\count204 by -\@p@sbblly
		\edef\ps@bbw{\number\count203}
		\edef\ps@bbh{\number\count204}
		\if@angle 
			\Sine{\@p@sangle}\Cosine{\@p@sangle}
	        	{\dimen100=\maxdimen\xdef\r@p@sbbllx{\number\dimen100}
					    \xdef\r@p@sbblly{\number\dimen100}
			                    \xdef\r@p@sbburx{-\number\dimen100}
					    \xdef\r@p@sbbury{-\number\dimen100}}
                        \def\minmaxtest{
			   \ifnum\number\p@intvaluex<\r@p@sbbllx
			      \xdef\r@p@sbbllx{\number\p@intvaluex}\fi
			   \ifnum\number\p@intvaluex>\r@p@sbburx
			      \xdef\r@p@sbburx{\number\p@intvaluex}\fi
			   \ifnum\number\p@intvaluey<\r@p@sbblly
			      \xdef\r@p@sbblly{\number\p@intvaluey}\fi
			   \ifnum\number\p@intvaluey>\r@p@sbbury
			      \xdef\r@p@sbbury{\number\p@intvaluey}\fi
			   }
			\rotate@{\@p@sbbllx}{\@p@sbblly}
			\minmaxtest
			\rotate@{\@p@sbbllx}{\@p@sbbury}
			\minmaxtest
			\rotate@{\@p@sbburx}{\@p@sbblly}
			\minmaxtest
			\rotate@{\@p@sbburx}{\@p@sbbury}
			\minmaxtest
			\edef\@p@sbbllx{\r@p@sbbllx}\edef\@p@sbblly{\r@p@sbblly}
			\edef\@p@sbburx{\r@p@sbburx}\edef\@p@sbbury{\r@p@sbbury}
		\fi
		\count203=\@p@sbburx
		\count204=\@p@sbbury
		\advance\count203 by -\@p@sbbllx
		\advance\count204 by -\@p@sbblly
		\edef\@bbw{\number\count203}
		\edef\@bbh{\number\count204}
}
\def\in@hundreds#1#2#3{\count240=#2 \count241=#3
		     \count100=\count240	% 100 is first digit #2/#3
		     \divide\count100 by \count241
		     \count101=\count100
		     \multiply\count101 by \count241
		     \advance\count240 by -\count101
		     \multiply\count240 by 10
		     \count101=\count240	%101 is second digit of #2/#3
		     \divide\count101 by \count241
		     \count102=\count101
		     \multiply\count102 by \count241
		     \advance\count240 by -\count102
		     \multiply\count240 by 10
		     \count102=\count240	% 102 is the third digit
		     \divide\count102 by \count241
		     \count200=#1\count205=0
		     \count201=\count200
			\multiply\count201 by \count100
		 	\advance\count205 by \count201
		     \count201=\count200
			\divide\count201 by 10
			\multiply\count201 by \count101
			\advance\count205 by \count201
		     \count201=\count200
			\divide\count201 by 100
			\multiply\count201 by \count102
			\advance\count205 by \count201
		     \edef\@result{\number\count205}
}
\def\compute@wfromh{
		\in@hundreds{\@p@sheight}{\@bbw}{\@bbh}
		\edef\@p@swidth{\@result}
}
\def\compute@hfromw{
	        \in@hundreds{\@p@swidth}{\@bbh}{\@bbw}
		\edef\@p@sheight{\@result}
}
\def\compute@handw{
		\if@height 
			\if@width
			\else
				\compute@wfromh
			\fi
		\else 
			\if@width
				\compute@hfromw
			\else
				\edef\@p@sheight{\@bbh}
				\edef\@p@swidth{\@bbw}
			\fi
		\fi
}
\def\compute@resv{
		\if@rheight \else \edef\@p@srheight{\@p@sheight} \fi
		\if@rwidth \else \edef\@p@srwidth{\@p@swidth} \fi
}
\def\compute@sizes{
	\compute@bb
	\if@scalefirst\if@angle
	\if@width
	   \in@hundreds{\@p@swidth}{\@bbw}{\ps@bbw}
	   \edef\@p@swidth{\@result}
	\fi
	\if@height
	   \in@hundreds{\@p@sheight}{\@bbh}{\ps@bbh}
	   \edef\@p@sheight{\@result}
	\fi
	\fi\fi
	\compute@handw
	\compute@resv}

\def\psfig#1{\vbox {
	\ps@init@parms
	\parse@ps@parms{#1}
	\compute@sizes
	\ifnum\@p@scost<\@psdraft{
		\special{ps::[begin] 	\@p@swidth \space \@p@sheight \space
				\@p@sbbllx \space \@p@sbblly \space
				\@p@sbburx \space \@p@sbbury \space
				startTexFig \space }
		\if@angle
			\special {ps:: \@p@sangle \space rotate \space} 
		\fi
		\if@clip{
			\if@verbose{
				\ps@typeout{(clip)}
			}\fi
			\special{ps:: doclip \space }
		}\fi
		\if@prologfile
		    \special{ps: plotfile \@prologfileval \space } \fi
		\if@decmpr{
			\if@verbose{
				\ps@typeout{psfig: including \@p@sfile.Z \space }
			}\fi
			\special{ps: plotfile "`zcat \@p@sfile.Z" \space }
		}\else{
			\if@verbose{
				\ps@typeout{psfig: including \@p@sfile \space }
			}\fi
			\special{ps: plotfile \@p@sfile \space }
		}\fi
		\if@postlogfile
		    \special{ps: plotfile \@postlogfileval \space } \fi
		\special{ps::[end] endTexFig \space }
		\vbox to \@p@srheight sp{
			\hbox to \@p@srwidth sp{
				\hss
			}
		\vss
		}
	}\else{
		\if@draftbox{		
			\hbox{\frame{\vbox to \@p@srheight sp{
			\vss
			\hbox to \@p@srwidth sp{ \hss \@p@sfile \hss }
			\vss
			}}}
		}\else{
			\vbox to \@p@srheight sp{
			\vss
			\hbox to \@p@srwidth sp{\hss}
			\vss
			}
		}\fi

	}\fi
}}
\psfigRestoreAt
\let\@=\LaTeXAtSign

\renewcommand{\topfraction}{0.9}                                                                                                                                                    
\renewcommand{\bottomfraction}{0.9}                                                                                                                                                 
\renewcommand{\textfraction}{0.05}                                                                                                                                                  
\renewcommand{\floatpagefraction}{0.05}                                                                                                                                             
\setcounter{topnumber}{5}                                                                                                                                                           
\setcounter{bottomnumber}{5}                                                                                                                                                        
\setcounter{totalnumber}{10}

\renewcommand\floatpagefraction{0.9}

\newcommand{\mb}[1]{\mbox{\boldmath$#1$}}

\newcommand{\aleq}{\mbox{\ 
\raisebox{-.9ex}{$\stackrel{\textstyle<}{\sim}$}\ }}
\newcommand{\ageq}{\mbox{\
\raisebox{-.9ex}{$\stackrel{\textstyle >}{\sim}$}\ }}

\topmargin -0.5in
\newcommand{\marg}[1]{\marginpar{\raggedright\scriptsize#1}}
\parskip=1pt
\setlength{\textwidth}{6.in}
\setlength{\textheight}{8.5in}
\setlength{\evensidemargin}{0.25in}
\setlength{\oddsidemargin}{0.25in}

\def\x{{\mbox{\boldmath$x$}}}
\def\u{{\mbox{\boldmath$u$}}}

\def\la{{\langle}}
\def\ra{{\rangle}}
\def\om{{\omega}}
\def\omd{{\omega_d}}
\def\nab{{\bf \nabla}}
\def\eps{{\epsilon}}
\def\begineq{\begin{equation}}
\def\endeq{\end{equation}}

\begin{document}
\bibliographystyle{prsty}

\title{
Sound radiation of 3\,MHz driven gas bubbles}
\author{Siegfried Grossmann, Sascha Hilgenfeldt, Detlef Lohse}
\address{
Fachbereich Physik der Universit\"at Marburg,
Renthof 6, 35032 Marburg, Germany}
\author{Michael Zomack}
\address{Schering AG, Clinical Development, M\"ullerstr.\ 178,
13342 Berlin, Germany}

\date{\today}

\maketitle

\vspace{1cm}

%%\centerline{{\em short title:} ``Sound radiation of driven bubbles''}
%\centerline{{\em footline text:} Grossmann {\em et al.}}

%\newpage

\begin{abstract}
The sound radiation of 3\,MHz acoustically driven air bubbles in
liquid is analysed with respect to possible applications in second harmonic
ultrasound diagnostics devices, which have recently come into clinical use.
In the forcing pressure amplitude $P_a = 1 - 10$\,atm and ambient radius
$R_0 = 0.5 - 5\,\mu m$ parameter domain a narrow regime around
the resonance radius $R_0 \sim 1- 1.5\,\mu m$ and relatively modest
$P_a \sim 2-2.5$\,atm is identified in which optimal sound yield in the second
harmonic is achieved while maintaining spherical stability of the bubble.
%We analyse the sound radiation of 3\,MHz acoustically driven air bubbles in
%liquid with respect to possible applications in second harmonic
%ultrasound diagnostics devices, which have recently come into clinical use.
%In the forcing pressure amplitude $P_a = 1 - 10$\,atm and ambient radius
%$R_0 = 0.5 - 5\,\mu m$ parameter domain we identify a narrow regime around
%the resonance radius $R_0 \sim 1- 1.5\,\mu m$ and relatively modest
%$P_a \sim 2-2.5$\,atm to achieve optimal sound yield in the second
%harmonic while maintaining spherical stability of the bubble.
For smaller $P_a$ and larger $R_0$ hardly any sound is radiated; for
larger $P_a$
bubbles become unstable towards
non-spherical shape oscillations of their surface. The computation
of these instabilities is essential for the evaluation of the
optimal parameter regime. 
A region of slightly smaller $R_0$ and $P_a\sim1-3$\,atm is best suited to
achieve large ratios of the second harmonic to the fundamental intensity.
Spherical stability is guaranteed in the suggested regimes
for liquids with an enhanced viscosity compared to water, such as blood.

%\vspace{1cm}

%\noindent
%PACS numbers: 87.59.Mt, 43.80.+p, 43.25.+y, 43.35.+d 
\end{abstract}

%\pacs{PACS numbers: 87.59.Mt, 43.80.+p, 43.25.+y, 43.35.+d} 

%\newpage

%----------------------------------------------------------------------
\section{Introduction}\label{introsec}
Microbubbles, i.e., gas bubbles of a few $\mu m$ diameter, have long
been known to be very effective scatterers of ultrasound
(cf.\ e.g.\ \cite{gra68}).  Their scattering cross sections for MHz sound
waves can be more than two orders of magnitude greater than their geometrical
cross sections \cite{nis75}. In the last decade,
the concept of exploiting this property to perform refined ultrasound
diagnostics with gas bubbles as echo contrast enhancers
has enjoyed increasing attention \cite{nan93}.
The general idea of this technique is to inject a microbubble
suspension into a vein and
to study the blood flow by detecting the bubbles' sound echo reaction to
an applied acoustical field. This leads to ultrasound images of higher
contrast
and quality as compared to conventional diagnostic techniques using only
the ultrasound backscatter from the tissue itself.

The quality of an ultrasonogram mainly depends on its spatial
resolution and its signal intensity, more specifically, on the
ratio of signal intensities from ``desired'' echoes (such as the
blood flow in bubble diagnostics) to ``background noise'' reflections
(from surrounding tissue). Spatial resolution is, of course, limited by the
wavelength of the
ultrasound. But as the absorption of sound in tissue increases exponentially 
with frequency \cite{seh82,jon86}, frequencies below 10\,MHz are used in
most clinical applications. As a typical value, we will choose an ultrasound
driving frequency of $\omd/2\pi=$3\,MHz throughout this work.

In doing diagnostics with bubble suspensions, it is desirable to
improve the signal to noise ratio. Namely, when detecting the emitted
sound from the bubble at the driving frequency 3\,MHz, the signal is
obscured by the driving and its reflections from tissue. 
To improve the signal quality, it has recently been proposed \cite{sch93,bur96}
 to detect {\it higher
harmonics} of the driving frequency in the sound emission spectrum of the
bubble. In view of the aforementioned strong damping of higher frequencies,
the lowest (second) harmonic at 6\,MHz is of particular interest. 
%One aim of this work is to show that the second harmonics
It can be
selectively excited if the parameters are chosen appropriately.
We expect that soft tissue, driven into the regime of nonlinear response by
the strong driving, will also reflect part of the sound in higher
harmonics. Also, the large amplitude driving signal itself will undergo
nonlinear distortion, generating higher harmonic frequency components
\cite{war97}.
As in the case of the conventional method, it may therefore be necessary to
focus on the
{\it difference} between the reflected signal with and without injected
microbubbles. However, by choosing the proper size of the bubbles and the proper
forcing pressure amplitude, it is possible to enhance the bubbles' reflection
signal in higher harmonics.
The main focus of this paper is to suggest a parameter regime
well suited for such an endeavor.

%%More references such as Church, etc...., prelim.\ studies...

Experimental and theoretical research on bubble dynamics has received
considerable attention since the discovery of single bubble
sonoluminescence by Gaitan
in 1990 (cf.\ \cite{gai92}) and the detailed experiments by the
Putterman group at UCLA \cite{bar91,hil94,put95}.
In those experiments single microbubbles are driven with a frequency of 
$\sim$ 30\,kHz and with a pressure amplitude of $P_a = 1.1 - 1.5$\,atm.
Under very special
conditions on experimental parameters such as the gas concentration in the
liquid and the pressure amplitude, 
the emission of short light pulses (once per driving period) from the center
of the bubble is observed. 
These experiments stimulated
us to perform a series of studies on bubble stability
\cite{bre95,bre96,hil96,loh96,bre96d,hil97}.
Three types of instability mechanisms seem to be important:
Spherical instability, diffusive instability, and chemical instability.
All of these studies are 
based on a Rayleigh-Plesset-like (RP) equation
%\cite{ray17,ple49,pro77,bre95b}
which provides an accurate description of the bubble wall dynamics even for
strongly nonlinear oscillations. Excellent 
agreement with the experiments was obtained, encouraging us to
rely on the RP equation also for microbubbles driven at 3\,MHz.
We will give an overview on RP dynamics in section \ref{rpsec}. Section
\ref{soundsec} shows results of calculated sound intensities emitted
into the whole spectrum as well as at the fundamental and second harmonic
frequencies. Of the instability mechanisms mentioned above, only shape
instabilities are important here.
They will be treated in section \ref{stabsec} and reveal important restrictions
on useful values of driving pressure amplitudes and bubble radii. Section
\ref{conclsec} presents conclusions. 

\section{Rayleigh-Plesset bubble dynamics}\label{rpsec}
We will treat the dynamics and sound emission of a single bubble here; we
assume that it is driven by a spatially homogeneous, standing wave field
\begineq
P(t) = P_a \cos \omd t
\label{eq1}
\endeq
with a frequency $\omd / 2\pi = 3$MHz, corresponding to a period
of $T=0.33\,\mu s$,  and a sound amplitude $P_a$ between $1$
and $10$\,atm (roughly $10^5$ to $10^6$Pa), reflecting typical peak pressures
of devices in ultrasound diagnostics. Much higher pressure amplitudes,
as applied in lithotripters (cf.\ e.g.\ \cite{chu89}), could damage the tissue.

Let us briefly discuss the approximations we made here.
In water, the chosen frequency corresponds
%\marg{subscript $c_{el}$}
to a sound wave length of $\lambda=2\pi c_l/\omd \approx 500\,\mu m$, where
$c_l=1481m/s$ is the sound velocity in water.
The typical ambient radius $R_0$ of the microbubbles is in the range
of $1-5\,\mu m$. Therefore, the approximation of spatial homogeneity
is justified. In diagnostic ultrasound devices, the driving sound is
not a standing wave, but a short traveling wave pulse (which is not
strictly monochromatic). The corresponding
spatial fluctuations of the driving pressure gradient
at the location of the bubble will exert translational forces on the
bubble. The resulting translational movements of the bubble are
neglected (we will come back to this
assumption later in this section) as well as bubble-bubble interactions,
the so-called secondary Bjerknes forces \cite{bre95b}.
Finally, pressure
fluctuations due to the blood pressure (order of magnitude
$0.05$\,atm) can also safely be neglected.
In many cases, ultrasound contrast enhancers do not contain pure air
bubbles, but stabilized bubbles with an albumin or saccharide coating
\cite{nan93}.
This may lead to a shift in the resonance frequency of the bubbles
as elaborated by de Jong, Church, and others \cite{dej91,dej92,chu95,ye96}.

Under these assumptions
the dynamics of the bubble radius $R(t)$ may be described by the
following ordinary differential equation
%Rayleigh-Plesset (RP) equation
\cite{ray17,bre95b}:
\begin{eqnarray}
R \ddot R + {3\over 2} \dot R^2  &=&
{1\over \rho_l} \left(p(R,t) - P(t) - P_0 \right)
             \nonumber \\
	            &+& {R\over \rho_l c_l} {d\over dt}
		    \left( p(R,t) - P(t)\right) - 4 \nu 
{\dot R \over R} -
		    {2\sigma \over
		    \rho_l R}.
		    \label{rp}
\end{eqnarray}
Typical parameters for an air bubble in water are
the surface tension
$\sigma = 0.073\,kg/s^2$, the water viscosity 
$\nu = 10^{-6}\,m^2/s$ and density 
$\rho_l= 1000\,kg/m^3$.
%\marg{subscript $\rho_{el}$}
We use these parameters for our calculations. Only the viscosity
is chosen to be larger by a factor of three with respect to water
($\nu = 3\cdot 10^{-6}\,m^2/s$), corresponding to the value for blood.
We will later see that the increased viscosity is
essential for the spherical stability of the bubble at higher values of $P_a$.
The external (ambient) pressure 
%\marg{subscript $P_{zero}$}
is $P_0= 1$\,atm. 
We assume that 
the pressure inside the bubble
is given by a van der Waals type equation of state
\begineq
p(R(t)) = \left(P_0 + {2\sigma \over R_0}\right)
\left( {R_0^3 - h^3\over R^3(t) - h^3}\right)^\kappa
\label{pressure}
\endeq
with a (collective) van der Waals hard core radius
$h= R_0/8.54$ (for air) \cite{loe93},
i.e., $h^3$ is a measure for the total excluded volume of the
molecules.
The bubble radius under normal conditions (ambient radius) $R_0$  is
not uniform for the bubble population in a diagnostic
suspension. The size distribution can, however, be controlled experimentally
and is typically centered around 1--2$\,\mu m$, with a width of about 
$1\,\mu m$ \cite{nan93}.
For the effective polytropic 
exponent we take $\kappa \approx 1$ as for the oscillation frequencies under
consideration
micrometer bubbles can be treated as approximately isothermal \cite{ple77}.
Equation (\ref{rp}) can be understood as a balance equation between the
excitation
due to the forcing (\ref{eq1}) on the one hand and dissipative and acoustic
loss processes on the other hand.

In this paper, we will denote (\ref{rp}) the (modified)
Rayleigh-Plesset (RP)
equation, adopting a common practice in recent work on sonoluminescence
\cite{bar92,loe93,loe95}. Besides the pioneering work of Lord Rayleigh
\cite{ray17} and Plesset \cite{ple49}, other researchers have contributed
to (\ref{rp}) and a number of variations of this equation, e.g.\
Keller and Miksis \cite{kel80}, Flynn \cite{fly75}, or Gilmore
\cite{gil52}. Some of these variations are much more elaborate
than (\ref{rp}). However, a detailed comparison of the solutions
obtained from these equations \cite{las81,hil97} shows that
significant deviations only occur for bubbles driven at very large
pressure amplitudes ($\ageq 5$\,atm) {\em and} having large radii, i.e.,
$R_0$ would have to be substantially larger than for the bubbles of
the present study to necessitate the use of a more complicated
dynamical equation.

%caption1
\begin{figure}[htb]
\setlength{\unitlength}{1.0cm}
\begin{center}
\begin{picture}(10,7.5)
\put(0.,0.){\psfig{figure=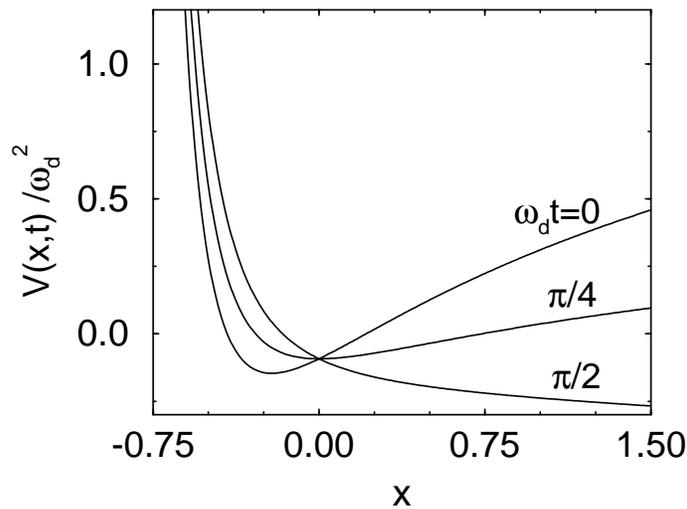,width=9.5cm,angle=-90.}}
\end{picture}
\end{center}
\caption[]{
Potential according to eq.\ (\ref{eq4}), non-dimensionalized through dividing
by $\omd^2$,
using $P_a=2$\,atm and $R_0=1.2\,\mu m$ for three
different phases $\omd t = 0,\, \pi/4,\, \pi/2$, respectively.
}
\label{fig_pot}
\end{figure}

Let us first consider the resonance structure of the RP oscillator in the
small forcing limit.
To calculate the main resonance frequency we first note that
for small forcing $P_a < P_0$ the contribution of the sound coupling to
the bubble dynamics (the term $\propto 1/c_l$
on the rhs of equation (\ref{rp})) is not important. In addition, if $R$
stays large enough to ensure $R^3 \gg h^3$ (which is the case for weak
driving), we can replace the van der Waals formula by an ideal
gas expression.
Writing $R(t) = R_0 (1 + x(t))$ we obtain 
\begin{eqnarray}
\ddot x & = & {1\over \rho_l R_0^2 }
\left[
\left( P_0 + {2\sigma \over R_0 } \right)
(1+x )^{-3\kappa-1 }- {P_0 + P_a \cos \omd t \over 1 + x }
-{2\sigma \over R_0 (1+x)^2} 
\right]  \nonumber \\
& - & {4\nu \dot x \over R_0^2 (1+x)} - {3\over 2}{\dot x^2 \over 1+x} .
\label{eq3}
\end{eqnarray}
We interpret the bracketed term on the rhs of eq.\ (\ref{eq3})
as an effective, time dependent
 force
$-\partial_x V(x,t)$.
Integration gives the time dependent potential
\begineq
V(x,t)= {1\over \rho_l R_0^2 }
\left[{1\over 3\kappa}\left( P_0 + {2\sigma \over R_0 } \right)
(1+x )^{-3\kappa  } + (P_0 + P_a \cos \omd t )\ln (1+x) - {2\sigma \over R_0 }
{1\over 1+x}\right] \; ,
\label{pot}
\endeq
which displays a strong asymmetry in $x$, see Fig.~\ref{fig_pot}.
If $P_a = 0$, the equilibrium point is $x=0$.
For general $P_a$ the minimum of the potential oscillates around this value.
If $P_a > P_0$ the potential is {\it repulsive} for a certain fraction of
the period.
For small $P_a$ and thus small $x$ we can linearize around $x=0$
and obtain a driven harmonic oscillator
\begineq
\ddot x + 2\gamma \dot x + \omega_0^2 x = {P_a\over \rho_l R_0^2} \cos \omd t
\label{eq4}
\endeq
%\marg{subscript $\omega_{zero}$}
with the eigenfrequency 
\begineq
\omega_0=\sqrt{{1 \over \rho_l R_0^2 }
\left(3\kappa P_0 + (3\kappa -1){2 \sigma \over R_0 }\right)}
\label{eq5}
\endeq
and the damping constant
$\gamma = {2\nu / R_0^2 }$.
For fixed driving frequency $\omd=2\pi\cdot 3$MHz and
$\kappa = 1$ the bubble oscillator is in (main) resonance if
$\omd=\omega_0$.
According to eq.\ (\ref{eq5}), this corresponds to a resonance radius of
$R_0 = 1.23\,\mu m$.
Taking viscous damping into account,
the oscillation amplitude has its maximum value at
a frequency \cite{ll60}
\begineq
\om^{(1)}_{res} = \sqrt{ \omega_0^2 - 2\gamma^2 } =
\sqrt{{1 \over \rho_l R_0^2 }
\left(3\kappa P_0 + (3\kappa -1){2 \sigma \over R_0 }\right)
- {8\nu^2\over R_0^4}},
\label{eq7}
\endeq
which (with $\nu = 3\cdot 10^{-6} m^2/s$) shifts the main
 resonance radius to
$R_0^{(1)} = 1.18\,\mu m$.  The corresponding radii for the subharmonics
$\om^{(1/2)}_{res} =\omd/2$,  
$\om^{(1/3)}_{res} =\omd/3$, and
$\om^{(1/4)}_{res} =\omd/4$ are
$R_0^{(1/2)} = 2.18\,\mu m$,  $R_0^{(1/3)} = 3.14\,\mu m$,
and $R_0^{(1/4)} = 4.09\,\mu m$, respectively.  The harmonic $\om^{(2)}_{res} = 2\omd$ is
at $R_0^{(2)} = 0.64\,\mu m$.
As we will see below, the harmonic and subharmonic resonances will
also strongly affect the intensity of the emitted sound.

%caption1
\begin{figure}[htb]
\setlength{\unitlength}{1.0cm}
\begin{center}
\begin{picture}(10,8.5)
\put(0.,0.){\psfig{figure=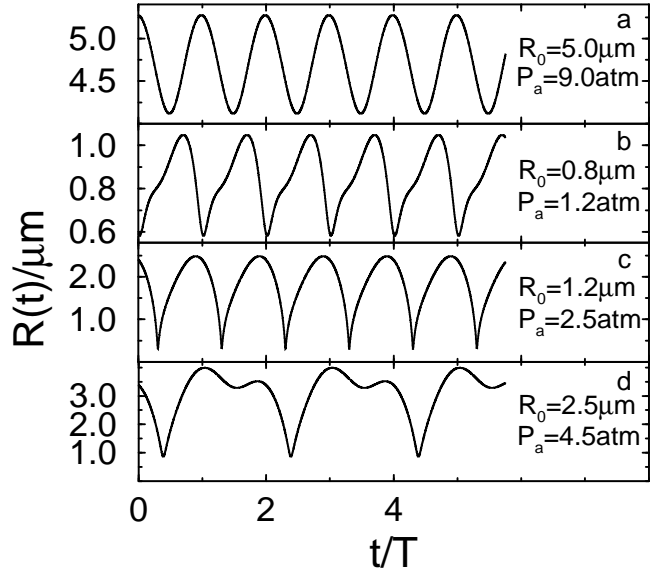,width=9cm,angle=0.}}
%\begin{picture}(12,12)
%\put(-0.7,-0.4){\psfig{figure=disk_09a:[sascha.tex.da.figs]schfig2da.eps,width=13.8cm,angle=0.}}
\end{picture}
\end{center}
\caption[]{
The bubble radius $R$ vs. time for four different parameter pairs ($R_0, P_a$);
from upper to lower: a) ($5\,\mu m$, $9.0$\,atm),
b) ($0.8\,\mu m$, $1.2$\,atm),
c) ($1.2\,\mu m$, $2.5$\,atm),
d) ($2.5\,\mu m$, $4.5$\,atm)
}
\label{fig_roft}
\end{figure}

Figure \ref{fig_roft}
shows the time series $R(t)$ for four typical sets of parameters, as
computed from (\ref{rp}) and (\ref{pressure})
using a double precision, fourth-order, variable stepsize Runge-Kutta
algorithm \cite{pre92}.
For small $P_a$
it is trivial that the radius oscillates sinusoidally, but this can also happen
for driving pressure amplitudes as large as $P_a= 9$\,atm, if the radius is
much larger than the resonance
radius, as seen in figure~\ref{fig_roft}\,a). For other parameter
combinations, the bubble
changes its behavior and exhibits {\it collapses}, characterized by
short duration minima of the radius, accompanied by large accelerations
(large curvature of $R(t)$).
In figure~\ref{fig_roft}\,b) we observe one (weak) bubble  
collapse per cycle which becomes stronger for larger $P_a$, see
figure~\ref{fig_roft}\,c).
For strong collapses the typical return time of the collapsing bubble
is in the nanosecond range.
Like the time series of most nonlinear oscillators,
the bubble dynamics can period double so that a collapse
only repeats
every two cycles, as shown in figure~\ref{fig_roft}\,d) for a strong collapse.
In other parameter regions, aperiodic behavior (``chaos'') can be
observed (cf.\ \cite{lau88}).
It can also be seen from Fig.~\ref{fig_roft} that our notion of
strong collapse coincides well with Flynn's \cite{fly75,fly88} 
definition of ``transient cavities'', for which the ratio of maximum
expansion radius and ambient radius (expansion ratio)
must fulfill $R_{max}/R_0 \ageq 2$: the examples of
Figs.~\ref{fig_roft}\,c and d show rapid collapses and an expansion ratio
of $\sim 2$.
On the other hand, Figs.~\ref{fig_roft}\,a and b
exemplify weakly oscillating bubbles with expansion ratios near one.

\begin{figure}[htb]
\setlength{\unitlength}{1.0cm}
\begin{center}
\begin{picture}(10,9.5)
\put(-2.8,0){\psfig{figure=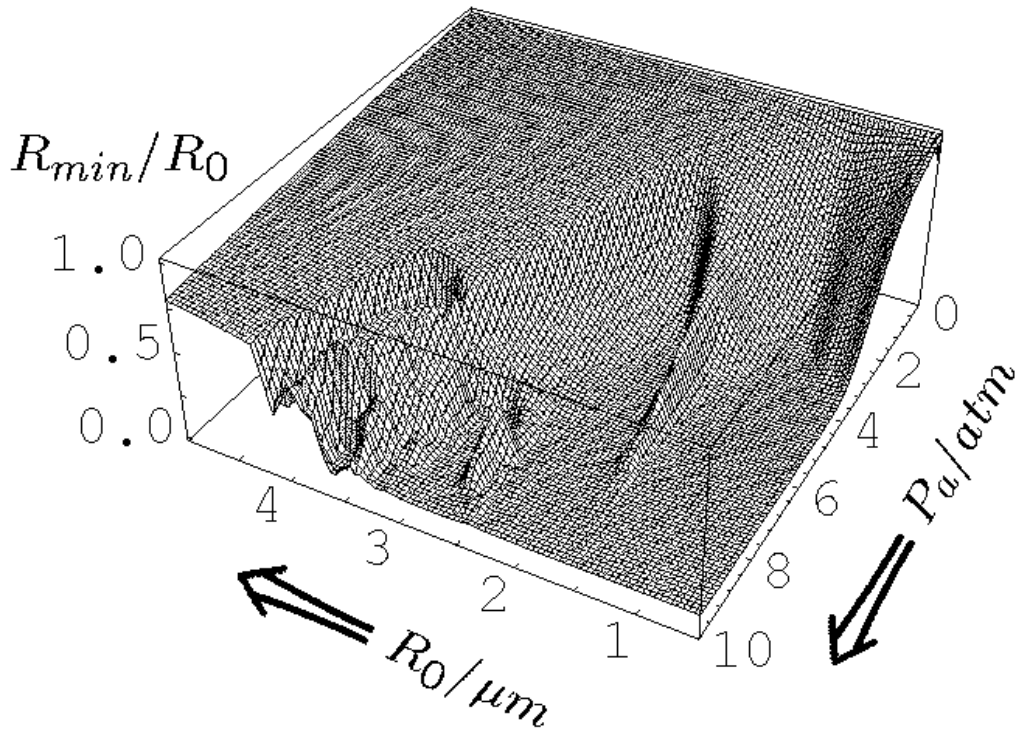,width=16cm,angle=0.}}
%\begin{picture}(12,11.5)
%\put(-2.8,-7.5){\psfig{figure=fig3.ps,width=18.5cm,angle=0.}}
\end{picture}
\end{center}
\caption[]{
Minimum radius $R_{min}/R_0$ as a function of $R_0$ and $P_a$. Note that the
graph is enclosed between two planes: for small $P_a$, the quotient
$R_{min}/R_0$ is nearly equal to one (weak oscillations),
for large $P_a$ and small
$R_0$ it approaches $h/R_0=1/8.54$, with the van der Waals hard core
radius $h$. Arrows in figures \ref{fig_rmin}, \ref{fig_total}, \ref{fig_2nd} --
 \ref{fig_1st2nd} indicate
axis orientation.
}
\label{fig_rmin}
\end{figure}

For the purposes of this paper, it is instructive to compute  
the {\it minimum} radius $R_{min} = \min_t (R(t))$ which the bubble achieves
during its oscillation. This quantity is shown in Fig.~\ref{fig_rmin}.
For this figure, as for the other 3D plots,
the displayed function was evaluated at $100\times 100$ equidistant
grid points in the $P_a - R_0$ plane.
For weak forcing the minimum radius essentially equals the ambient
radius (limit of small oscillations). However, if the driving pressure
amplitude is large enough, the bubble collapse is only halted in the
immediate vicinity of the smallest possible radius, i.e., the van der Waals
hard core radius. 
The transition towards hitting the hard core radius
$h=R_0/8.54$ (upon increasing $P_a$) is rather abrupt, forming a
well-defined threshold in the $P_a$ -- $R_0$ plane.
Obviously, this transition occurs for smaller
$P_a$ if the bubble radius is near one of the above mentioned resonance
radii.  
The resonances at $R_0^{(1)} = 1.18\,\mu m$ and $R_0^{(1/2)} = 2.18\,\mu m$
are clearly recognized in Fig.~\ref{fig_rmin}. As $P_a$ becomes larger,
the nonlinearities in (\ref{rp}) lead to broadening and a slight shift
of the resonances towards smaller $R_0$, in
accordance with earlier work on bubble dynamics \cite{ple77,kel80}.

%caption1

We now come back to our above assumptions on the pressure field.
The results of this work were obtained assuming a driving by standing
plane waves.
Today's ultrasound diagnostics devices usually emit a bundle of
traveling waves that interfere constructively to build up large pressure
peaks ($5-10$\,atm) and to achieve sufficient spatial resolution.
The resulting spatial pressure gradients lead to
translational forces acting on the bubble, 
the so called primary Bjerknes forces \cite{bre95b}:
\begineq
\mb{F_B}(\mb{r},t)=V_b(t) \mb{\nabla} P(\mb{r},t)\, , 
\label{bjerknes}
\endeq
where $V_b(t)$ is the time dependent volume of the bubble,
and $P(\mb{r},t)$ is the external pressure exerted on the bubble.
However, numerical computation of (\ref{bjerknes}) shows
that the accelerations resulting from a pressure gradient
$|\mb{\nabla}P|\sim P_a/\lambda$
are rather weak and that the
translational velocities of the bubbles are small compared
to their radial oscillation velocities. 

%it takes
%thousands of cycles to move the bubble from an antinode to a node.
%As the applied sound fields in ultrasound diagnostics generally
%have pulse durations much shorter than this 
%time scale, we neglect the effect of Bjerknes forces.
%This
%focusing is, however, relatively weak and the resulting pressure
%gradients will not be much greater than those resulting from~(\ref{bjerknes}).

%These forces push bubbles smaller than the main resonance radius $R_0^{(1)}$
%towards a pressure antinode and bubbles larger than the resonance radius
%%(i.e., $R_0 > 1.18\,\mu m$ in our case)
%towards the pressure node where they would
%not experience any pressure fluctuations.

%where the angular brackets denote averaging over one period of driving,

\section{Sound emission of oscillatory bubbles}\label{soundsec}
Our focus of interest is on the sound emitted by the oscillating bubble.
The far field sound pressure at a distance $r\gg R$ from
the center of the bubble can be calculated as  \cite{bre95b,ll87}
\begineq
P_s(r,t) = {\rho_l \over 4\pi r} {d^2V_b \over dt^2} = \rho_l {R\over r}
\left( 2 \dot R^2 + R\ddot R\right).
\label{sound}
\endeq
An ultrasound diagnostics device will display a picture of sound intensity,
which is based on the modulus
(or, equivalently, the square) of the sound pressure. Obviously, the
total detected intensity will not consist exclusively of signals due to
(\ref{sound}), but there will also be intensity components from reflections
of the incoming signal (\ref{eq1}) in the tissue. As
these latter contributions depend on many peculiarities of the
experimental or diagnostic setup, we do not try to model them here, but
focus on the active sound radiation of the bubble.

\begin{figure}[htb]
\setlength{\unitlength}{1.0cm}
\begin{center}
%\begin{picture}(12,18)
%\put(0.0,17){\LARGE a)}
%\put(0.0,8.2){\LARGE b)}
%\put(-0.5,3.){\psfig{figure=fig4a.ps,width=15.2cm,angle=0.}}
%\put(-0.5,-6.6){\psfig{figure=fig4b.ps,width=15.2cm,angle=0.}}
%%\put(0.7,3.){\psfig{figure=dummy.eps,width=12cm}}
%%\put(0.7,-6.6){\psfig{figure=dummy.eps,width=12cm}}
\begin{picture}(12,17.5)
\put(-1.0,15){\LARGE a)}
\put(-1.0,7.5){\LARGE b)}
\put(0.,8.7){\psfig{figure=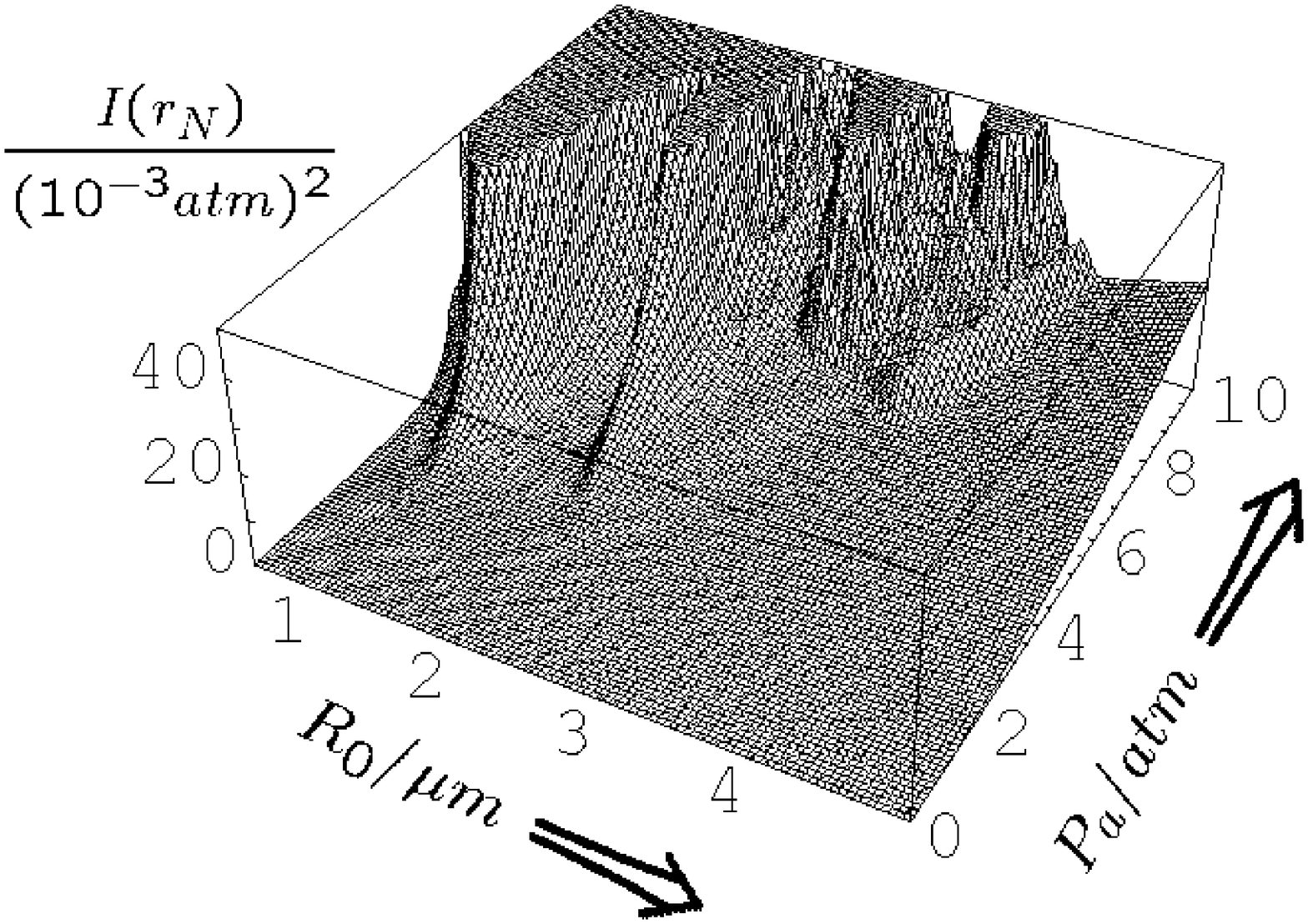,width=12cm,angle=0.}}
\put(0.,0.2){\psfig{figure=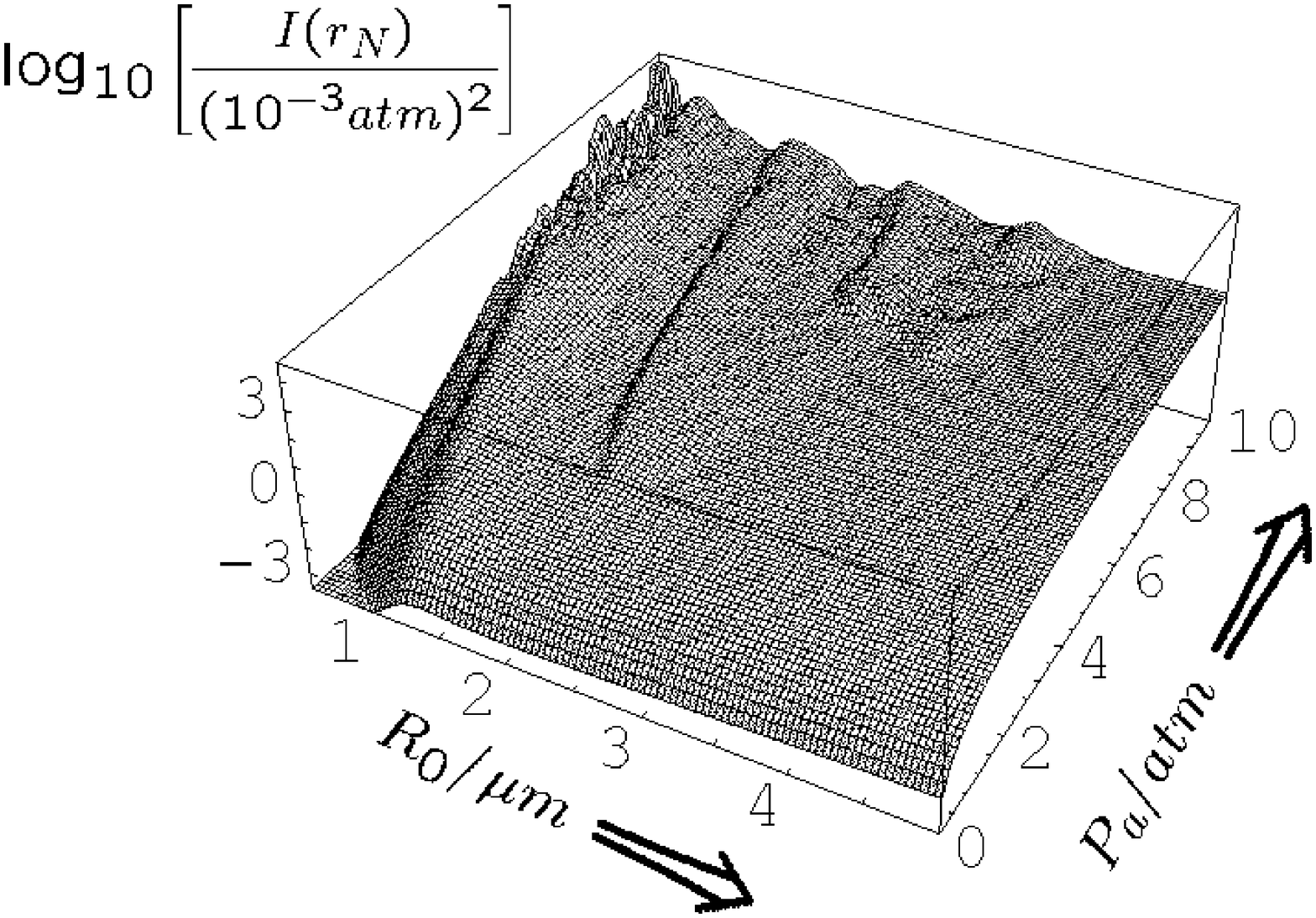,width=12cm,angle=0.}}
\end{picture}
\end{center}
\caption[]{
a)
Total sound intensity $I$ at a distance of $r_N=1$cm from the bubble center
as a function of $R_0$ and $P_a$. The strong correlation
between this figure and figure \ref{fig_rmin} is obvious.
Note that the graph is truncated (higher $I$-values are not displayed)
for better illustration of the resonance tongue structure. Figure
b) shows a logarithmic plot of the same quantity. The small undulations at
very large $P_a$ on top of the resonance structure are due to numerical
aliasing in the Fourier analysis and can be reduced with increasing computer
power.
}
\label{fig_total}
\end{figure}

Figure \ref{fig_total}a shows the {\it total} sound intensity $I$ 
as a function of the forcing pressure amplitude $P_a$ and the ambient radius
$R_0$.
According to Parseval's theorem it can be calculated either from the sound
pressure time series $P_s(r,t)$ or from its Fourier transform $P_s(r,\om)
= \int_0^\tau P_s(r,t) \exp(i\om t)dt$, $\tau \gg T$, namely
\begineq
I(r)={1\over \tau} \int_0^\tau |P_s (r,t) |^2 dt = {1\over \tau}
\cdot{1\over 2\pi}\int_{-\infty}^{+\infty} |P_s (r,\om)|^2 d\om.
\label{eq8}
\endeq
We divide by the length $\tau$ of the time series ($\tau\geq 8T$ for all
computations) and measure $I$ in units
of atm$^2$. The intensity is evaluated at a distance of $r_N=1$cm from the
bubble's center using standard double precision Fourier transform algorithms
\cite{pre92}. As $I$ spans several orders of magnitude in our parameter
regime, we also present its logarithm in Fig.~\ref{fig_total}\,b.

%caption1

It can be seen from Fig.~\ref{fig_total} that 
for small $P_a$ or large $R_0$ the bubble hardly emits any sound. Sound losses
set in at a sharp threshold which is very similar to
 the threshold seen in Fig.
\ref{fig_rmin} for the minimum radius.
Moreover, comparing Figs.~\ref{fig_rmin}
and \ref{fig_total} one realizes that strong sound emission
and collapsing to the hard core radius are strongly correlated (note the
opposite orientations of these two graphs). This
behavior is expected from equation
(\ref{sound}), as
a stronger collapse means larger bubble wall acceleration $\ddot R$
at the moment of collapse.
The resonance structure in $R_0$ is also
clearly reflected in the emitted sound intensity.

%caption1
\begin{figure}[thb]
\setlength{\unitlength}{1.0cm}
\begin{center}
%\begin{picture}(12,13.5)
%\put(-1.2,-0.4){\psfig{figure=disk_09a:[sascha.tex.da.figs]schfig5da.eps,width=13.3cm,angle=0.}}
\begin{picture}(10,9)
\put(0.,0.){\psfig{figure=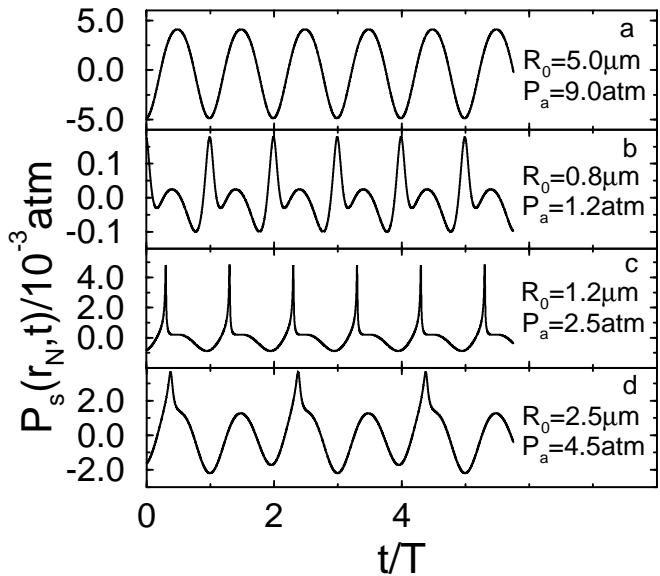,width=9cm,angle=0.}}
\end{picture}
\end{center}
\caption[]{
Time series of the sound pressure $P_s(r_N,t)$ from (\ref{sound})
for the same four pairs of parameters as in figure
\ref{fig_roft}.
}
\label{fig_sound}
\end{figure}

\begin{figure}[thb]
\setlength{\unitlength}{1.0cm}
\begin{center}
%\begin{picture}(12,11.5)
%\put(-0.5,-0.2){\psfig{figure=disk_09a:[sascha.tex.da.figs]schfig6da.eps,width=13.3cm,angle=0.}}
\begin{picture}(10,9)
\put(0.,0.){\psfig{figure=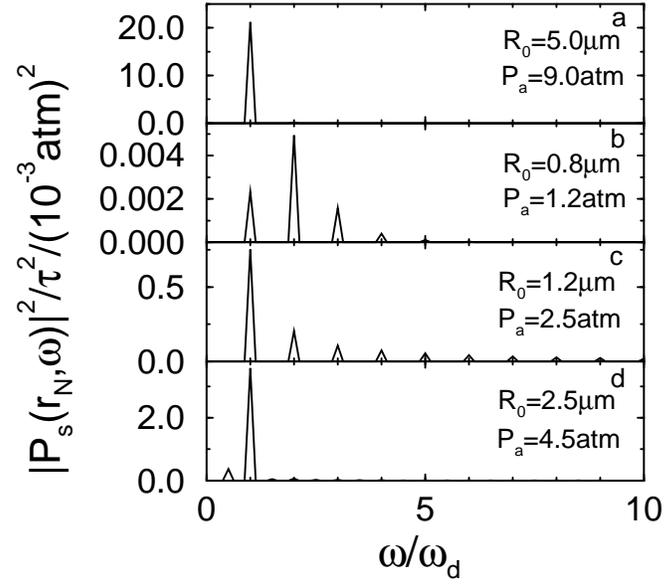,width=9cm,angle=0.}}
\end{picture}
\end{center}
\caption[]{
Power spectra $|P_s(r_N,\omd)|^2$
for the same four pairs of parameters as in figure
\ref{fig_roft}.
}
\label{fig_spectra}
\end{figure}

\begin{figure}[htb]
\setlength{\unitlength}{1.0cm}
\begin{center}
%\begin{picture}(12,10.6)
%\put(-1.8,-7.7){\psfig{figure=fig7.ps,width=18.1cm,angle=0.}}
\begin{picture}(10,9.5)
\put(-2.8,-1.){\psfig{figure=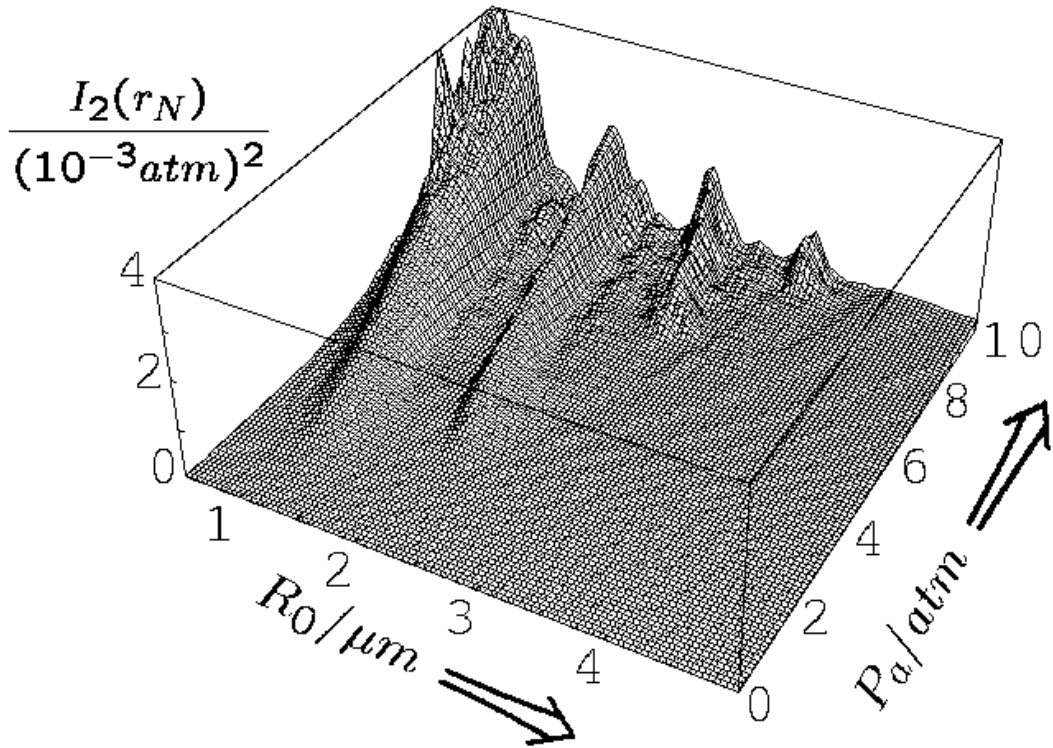,width=15cm,angle=0.}}
\end{picture}
\end{center}
\caption[]{
Absolute intensity of the second harmonic
of the sound emission (\ref{sound}).
}
\label{fig_2nd}
\end{figure}

\begin{figure}[htb]
\setlength{\unitlength}{1.0cm}
\begin{center}
%\begin{picture}(12,10.6)
%\put(-1.8,-7.7){\psfig{figure=fig8.ps,width=18.1cm,angle=0.}}
\begin{picture}(10,10)
\put(-2.8,0.){\psfig{figure=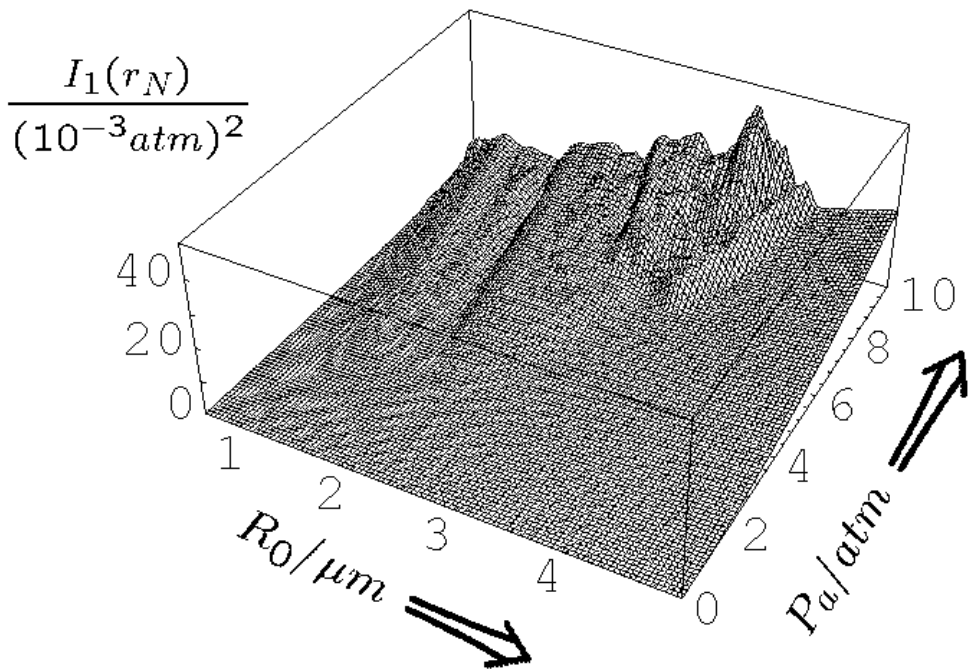,width=15cm,angle=0.}}
\end{picture}
\end{center}
\caption[]{
Absolute intensity of the fundamental
of the sound emission (\ref{sound}).
}
\label{fig_1st}
\end{figure}

Time series of $P_s(r_N,t)$ and their power spectra for the same four
parameter pairs ($R_0$, $P_a$) as in Fig.~\ref{fig_roft}
are shown in Fig.~\ref{fig_sound} and Fig.~\ref{fig_spectra}, respectively.
If the collapse is too violent,
the sound intensity is distributed in a broad band spectrum with a large
fraction of intensity emitted at higher harmonics, which will 
be absorbed by the tissue.
%(the absorption length is approximately inversely
%proportional to the frequency \cite{seh82,jon86}).
Therefore it is appropriate
to focus on the signal at the second harmonic frequency in order to get
high intensities away from the driving frequency. 
The intensity $I_2(r)=|P(r,\om = 2\omd)|^2/\tau^2$
in the second harmonic is displayed in Fig.~\ref{fig_2nd}.
%, whereas
%\ref {fig_2nd}b shows the fraction of the total spectral intensity emitted
%at this frequency.
For comparison, we show the same plot also for the 
%\marg{subscript $I_{one}$}
intensity $I_1(r)=|P(r,\om = \omd)|^2/\tau^2$ of the fundamental
(driving frequency) in Fig.~\ref{fig_1st}.
% and \ref{fig_1st}b, respectively.

%caption1

%caption1

Not surprisingly, the regions of greatest intensity of the fundamental
as well as of the second harmonic coincide with
those of the total intensity, i.e., they can be found at the resonance radii.
The largest region and the highest maxima of second harmonic intensity
occurs around $R_0^{(1)}$.
For small $P_a$ the intensity is of course nearly exclusively in the driving
frequency itself, which is the frequency of the small oscillations
of the bubbles. Upon
increasing $P_a$, sound is also emitted in the second harmonic mode, for
some parameter combinations up to 40\% of the total intensity. At even larger
$P_a$, higher and higher modes are excited, leaving a smaller and smaller
fraction of total intensity for the second mode (cf.\ Fig.~\ref{fig_spectra}\,c
and d).
%and \ref{fig_2nd}).
Even for large $P_a$, the driving frequency $\omd$ remains the largest
component of total emitted power in spite of the strong collapse, which
displays much larger peak pressures, but only lasts for extremely
short periods of time.

%caption1

%caption1
\begin{figure}[htb]
\setlength{\unitlength}{1.0cm}
\begin{center}
%\begin{picture}(12,11.)
%\put(-1.8,-7.7){\psfig{figure=fig9.ps,width=18.4cm,angle=0.}}
\begin{picture}(10,10)
\put(-2.8,-1.){\psfig{figure=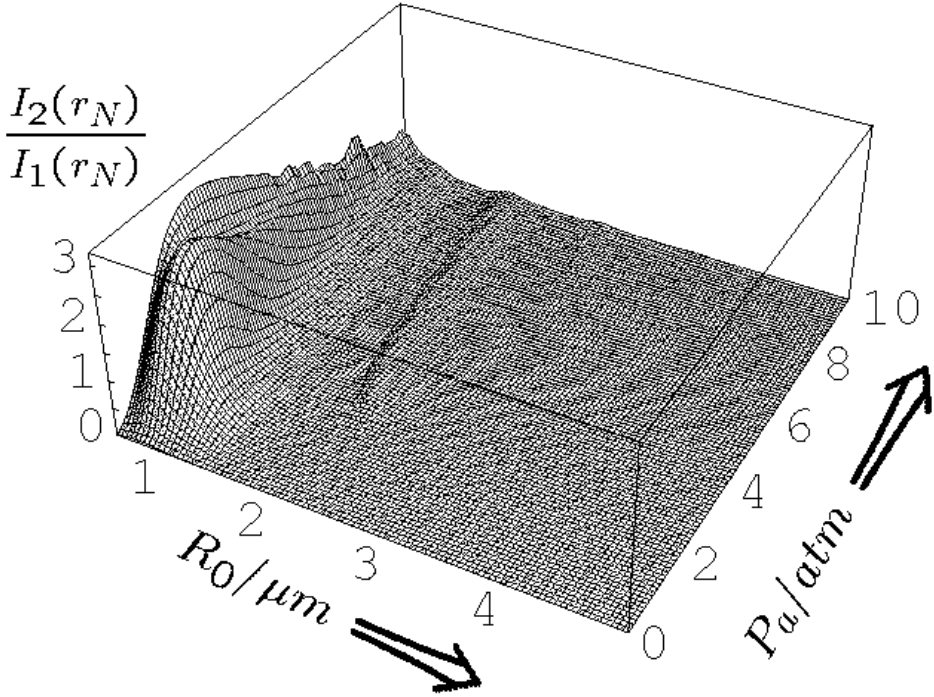,width=15cm,angle=0.}}
\end{picture}
\end{center}
\caption[]{
Ratio $I_2/I_1$ of intensities of sound emission at the second harmonic and
fundamental frequencies.
}
\label{fig_1st2nd}
\end{figure}

The key question now is: How should one choose the parameters for
optimal detection of the second harmonic? 
The answer cannot be given exclusively from
the absolute intensity of the second harmonic, Fig.~\ref{fig_2nd}.
Clearly, the signal has to have a certain absolute intensity to
overcome the noise level, and therefore Fig.~\ref{fig_2nd} gives
important information. One important
 application of the second harmonic method
in ultrasound diagnostics, however, relies on the contrast between
intensities at fundamental and second harmonic frequencies:
When injecting a bubble suspension into the vascular system, a second harmonic
signal is only
expected from the contrast agent. Thus, detecting at 6\,MHz in our example
will give a bright image of the blood vessels. 
In a diagnostic situation, it should be possible to {\it switch}
between this image and the scattering signal from surrounding tissue,
which reflects the 3\,MHz driving. But if the bubble emission signal at 3\,MHz
is more intense than these reflections, the vascular system will be
the dominant feature in the fundamental frequency image, too.
Therefore, meeting the demands of this application means to
identify parameter regions where the {\it ratio} of second harmonic
intensity to fundamental intensity $I_2/I_1$ is as high as possible.
Fig.~\ref{fig_1st2nd} displays this quantity.
It shows a distinct maximum at small $R_0 \approx 0.7-0.8\,\mu m$ close
to the harmonic bubble resonance radius $R_0^{(2)}$. In this parameter
region,
the bubble essentially oscillates with 6\,MHz instead
of 3\,MHz; an example for this behavior can be seen in Figs.~\ref{fig_sound}b and
\ref{fig_spectra}b.
For even smaller radii, a rise in $I_2/I_1$
indicates other resonances.
All bubbles with high $I_2/I_1$ ratios emit little absolute acoustic
intensity (cf.\ Figs.~\ref{fig_total} --  \ref{fig_1st}).

%In a region of low absolute intensity,
%the relative intensity of the second harmonic mode
%displays a distinct maximum at small $R_0 \approx 0.7\,\mu m$
%in the latter figure. This maximum 
%occurs when the eigenfrequency of the bubble is twice the forcing
%frequency $\omd$.

%Relative intensities become important
%when the incoming sound from the ultrasonic device consists of wavetrains
%only a few wavelengths long, which is often the case in diagnostic
%applications. Necessarily, in this case the bandwidth of the signal 
%is very large and the signal at 6\,MHz will contain components that are
%not due to active emission by the bubble, but passive reflections of
%the incoming waves. In this respect, the ratio of intensities at the
%second harmonic and the fundamental frequency is of primary importance.

We have presented a variety of intensity diagrams in Figs.~\ref{fig_total}
-- \ref{fig_1st2nd} in order to meet the different demands of different
experimental setups or diagnostic applications. Accordingly, the optimal
parameter ranges for second harmonic sonography depend on the focus of
interest:
If the important quantity is relative intensity $I_2/I_1$,
one would pick the maximum of Fig.~\ref{fig_1st2nd}, i.e., bubbles
with $R_0\approx 0.5 - 1.2\,\mu m$ and driving pressure amplitudes
$P_a\approx 1 - 3$\,atm.  
If absolute intensity is the key variable,
one would choose the region around the main resonance radius in $R_0$, i.e.,
$R_0\approx  1.0 - 1.5\,\mu m$. Moreover, Fig.~\ref{fig_2nd} suggests
choosing the pressure amplitude $P_a$ as large as possible. However,
as we will show in the next paragraph, bubble shape instabilities set an
upper limit on practically useful $P_a$. Note that a change in the driving
frequency would shift the resonance radii and, consequently, also the
location of regions of maximum sound emission. If, for example, $\omd$
is smaller, the resonances are shifted towards larger $R_0$, see 
equation (\ref{eq5}).

Also, the total amount of bubbles should be large enough to guarantee a
strong 
signal, but low enough to prevent considerable bubble-bubble interaction.
Therefore, it is of primary importance to assure that a high percentage of the
generated bubbles is in the correct $R_0$ regime.
All other bubbles are essentially useless regarding the yield in the second
harmonic and may even obscure the measurements.

\section{Spherical stability}\label{stabsec}
\begin{figure}[htb]
\setlength{\unitlength}{1.0cm}
%\begin{picture}(15,16.8)
%\put(-0.5,6.2){\psfig{figure=fig10a.ps,width=8.8cm,angle=0.}}
%\put(7.2,6.2){\psfig{figure=fig10b.ps,width=8.8cm,angle=0.}}
%\put(3.3,-2.2){\psfig{figure=fig10c.ps,width=8.8cm,angle=0.}}
\begin{picture}(15,17)
\put(-0.5,6.8){\psfig{figure=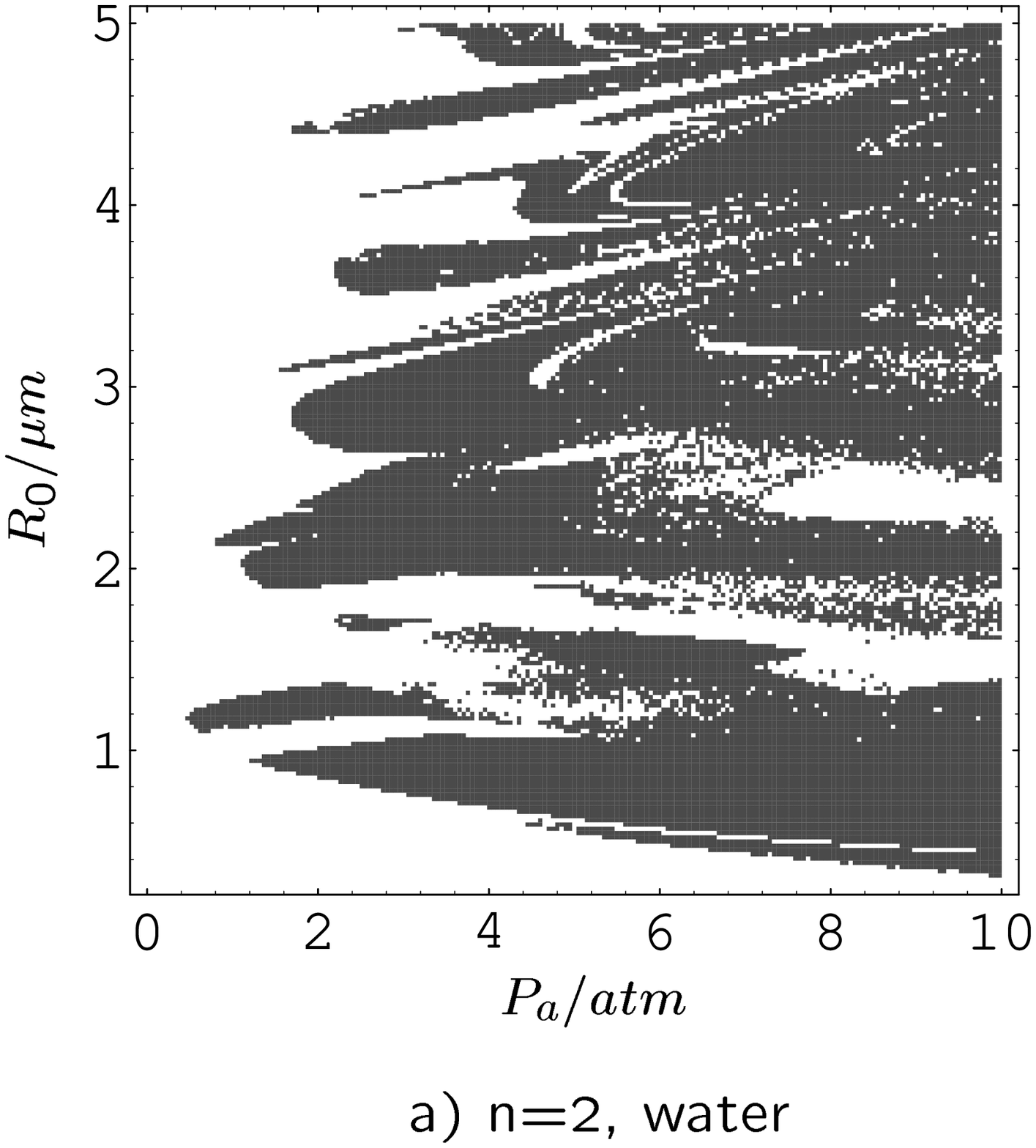,width=8.8cm,angle=0.}}
\put(7.2,6.8){\psfig{figure=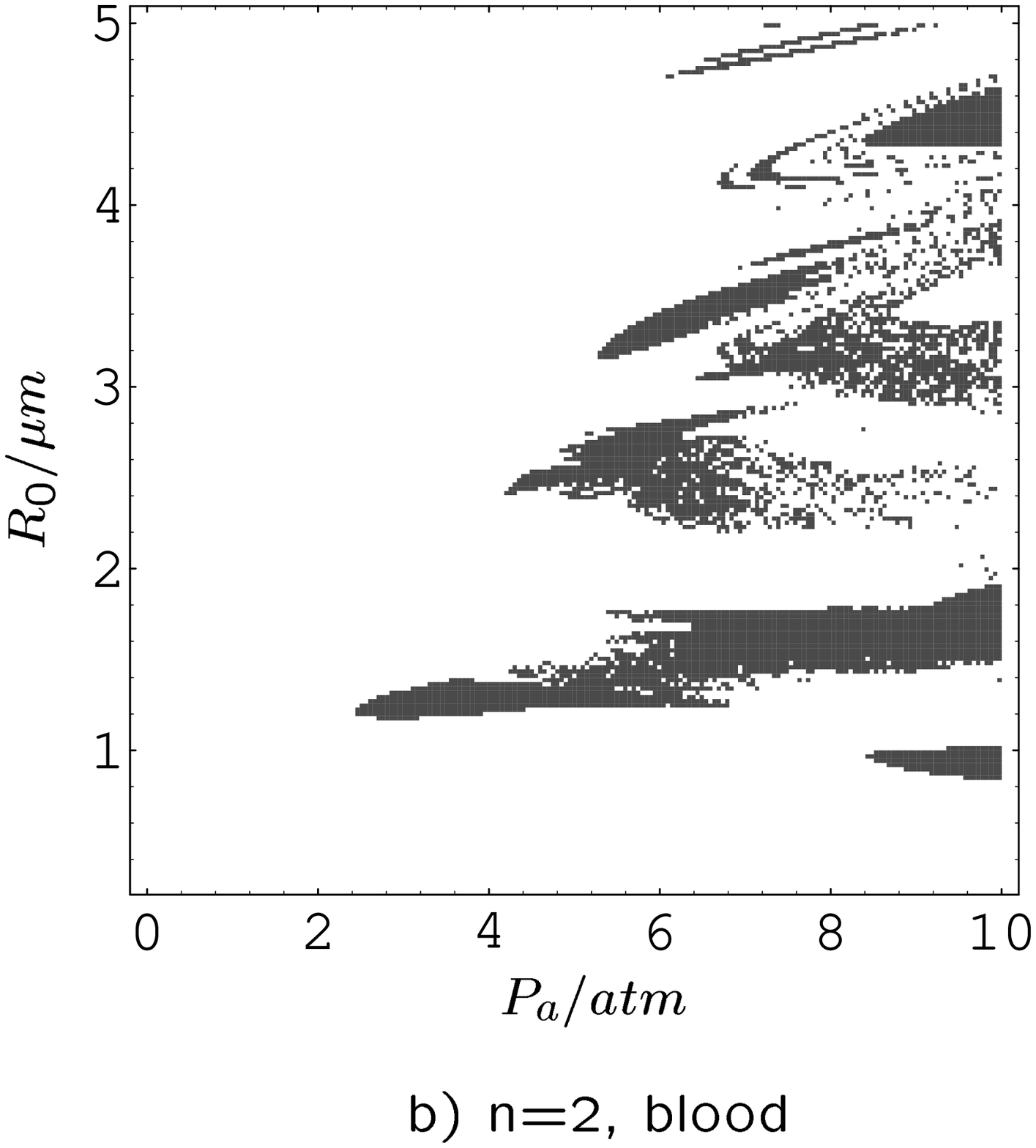,width=8.8cm,angle=0.}}
\put(3.3,-1.6){\psfig{figure=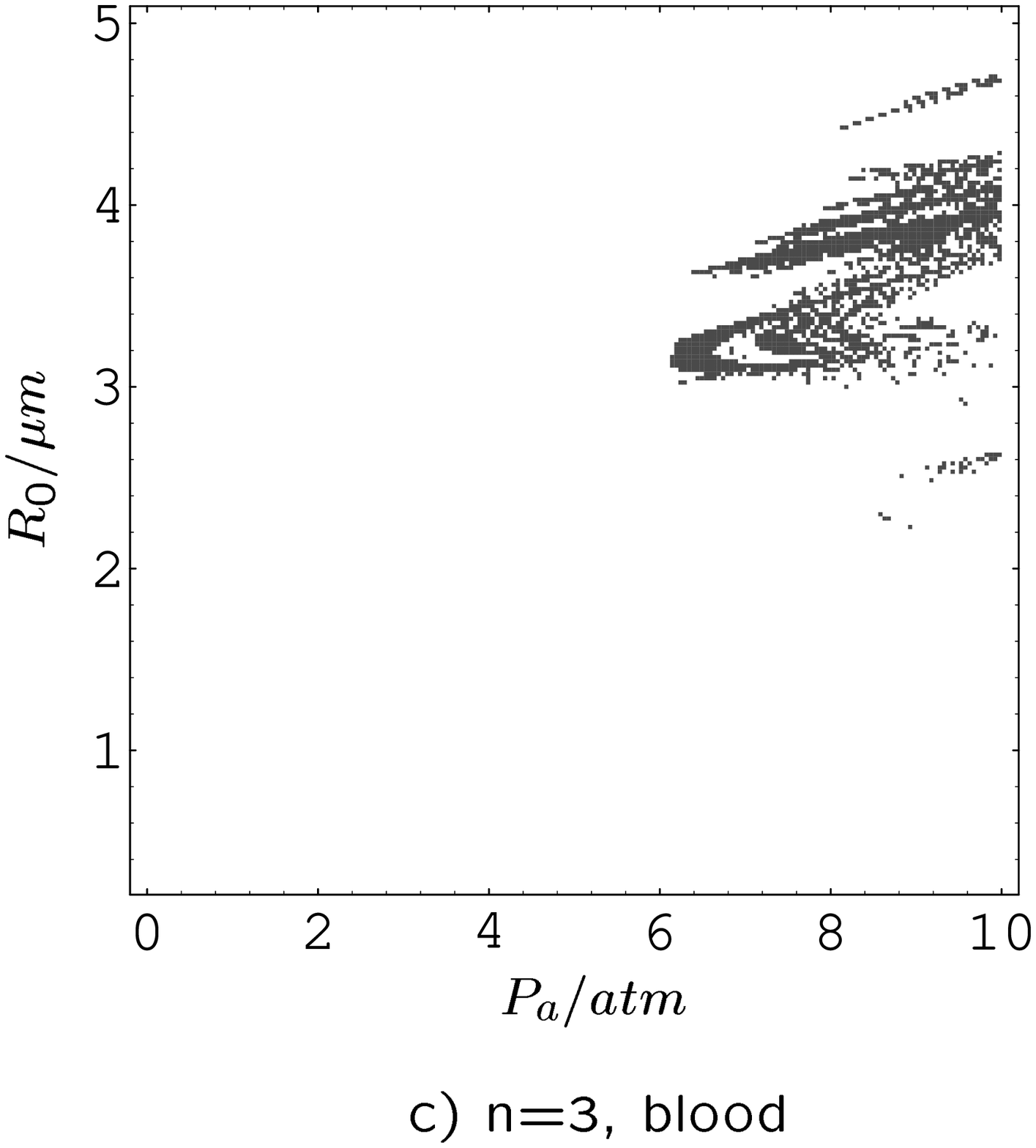,width=8.8cm,angle=0.}}
\end{picture}
\caption[]{
Stability diagram
for the $n=2$ mode for bubbles in water (a)
and in  blood (b)  and for the $n=3$ mode for bubbles in blood (c)
according to the Floquet
multipliers computed from equation (\ref{floquet}).
In the white regions
the bubbles are parametrically stable towards perturbations of
the corresponding mode,
in the dark regions they are parametrically
unstable.
Bubbles in water are much less stable than those in blood and
higher modes $a_n$ with
$n\ge 3$ tend to be more stable than the second mode $a_2$.
We stress that details of
these stability diagrams depend on our approximation (\ref{pro8})
-- (\ref{pro12}) as well as on the choice of the parameters of the liquid
and the averaging time which is $20T$ here.
Thus they should only be considered as  a reflection
of the general trend.
}
\label{fig_stab}
\end{figure}

To take advantage of the 
above suggested parameter regimes derived from RP dynamics, the bubbles
in these domains should be {\it spherically stable}, i.e., stable against
the growth of
non-spherical bubble deformations, which could eventually lead
to bubble fragmentation and a breakdown of sound emission. The
corresponding stability analysis has been
performed in detail in \cite{hil96}. We give a brief summary here. 
Consider a small distortion
of the spherical interface $R(t)$,
$$R(t) + a_n(t) Y_n(\theta , \phi ),$$
where $Y_n$ is a spherical (surface) harmonic of degree
$n$.  
An approximate
 linearized dynamical equation of the distortion $a_n(t)$ for each 
mode has been derived in \cite{hil96}, following the pioneering work
of Prosperetti \cite{pro77}. It reads (cf.\ \cite{bre95})
\begin{eqnarray}
\ddot a_n + B_n(t) \dot a_n - A_n(t) a_n=0
\label{pro8}
\end{eqnarray}
with
\begineq
A_n(t) = (n-1 ) {\ddot R \over R } - {\beta_n \sigma 
\over \rho_w R^3}
- {2 \nu \dot R \over R^3} \left[
(n-1)(n+2) + 2n (n+2) (n-1)
{\delta \over R }
\right] ,
\label{pro13}
\endeq
\begineq
B_n(t) = {3\dot R \over R } +  {2\nu \over R^2 }
\left[
(n+2) (2n+1) - 2n (n+2)^2 
 {\delta \over R}
 \right].
\label{pro14}
\endeq
Here,
$\beta_n = (n-1)(n+1)(n+2)$ and $\delta$ is
a viscous boundary layer cutoff \cite{hil96},
\begineq
\delta = min\left( \sqrt{{\nu \over \omd }}, {R\over 
2n} \right).
\label{pro12}
\endeq
%Our results are based on these equations.
If the coefficients $A_n(t)$ and $B_n(t)$ are periodic with period $T$,
(\ref{pro8}) is an equation of Hill's type and 
instability occurs whenever
the magnitude of the maximal eigenvalue of
the Floquet transition matrix $F_n(T)$ of eq.\ (\ref{pro8})
is larger than one. 
The Floquet transition matrix
$F_n(T)$ is defined by 
\begin{equation}
\left(\begin{array}{c}
a_n(T) \\ \dot a_n(T)
\end{array}\right)= F_n(T)
\left(\begin{array}{c}
a_n(0) \\ \dot a_n(0)
\end{array}\right).
\label{floquet}
\end{equation}
Now for some parameter regimes the radius is not periodic with period $T$
and thus
the coefficients $A_n(t)$ and $B_n(t)$ are not  either.
Therefore, rather than calculating the Floquet matrix $F_n(T)$ we
calculate a transition matrix $F_n(NT)$ with a large integer $N$
(here $N=20$) to average the
dynamics. We 
numerically compute the eigenvalues of $F_n (NT)$.
The logarithm of the maximum eigenvalue
can be understood as an approximate Lyapunov exponent.
If it is positive, the mode $a_n(t)$ grows exponentially
and the bubble is unstable towards the corresponding mode of
shape oscillation.
In Fig.~\ref{fig_stab}\,b and \ref{fig_stab}\,c
we show the resulting stability diagrams for
the second and the third mode ($n=2$ and $n=3$, respectively).
Bubbles are shape unstable in the dark regions of the $P_a - R_0$ plane,
and shape stable in the white areas.
Generally, the $n=2$ mode is the most unstable one, but there are
regimes where this does not hold.

The most pronounced features of these stability diagrams are
``tongues'' of instability.
In the low $P_a$ regime equation (\ref{pro8})
reduces to a Mathieu equation; in this case the tongues of instability
are the well known Mathieu tongues, as demonstrated in \cite{bre95}.
Large viscosity strongly damps out
this tongue structure, as seen from comparing the stability diagrams for
water and blood (different viscosities) in Fig.~\ref{fig_stab}a and
\ref{fig_stab}b.
In some regions of parameter space, stable and unstable points seem to be
mixed erratically. This is due to long-periodic or chaotic bubble dynamics
for these $P_a - R_0$ combinations, for which the results of
our stability analysis over $20T$ depend sensitively on the initial
conditions, so that for slightly deviating parameters the stability
behavior may be completely different. 
 
%caption1

As we learn from this figure, bubbles in the $R_0$ regime of maximum
(absolute) yield in the second harmonic become spherically unstable for
drivings
with $P_a\ageq2.5$\,atm. Below, the bubbles are stable in blood (due to its
enhanced viscosity),
whereas for water these bubbles are unstable
with respect to the $a_2$-mode even at $P_a\approx0.5$\,atm.
Note that it is risky even to be
close to an unstable regime, as the bubble may
diffusionally shrink
or grow into these regimes. 
This suggests that experiments with bubble suspensions in water will
probably give misleading results (with respect to clinical applications)
and should rather be carried out in blood or a fluid of correspondingly
enhanced viscosity. In our calculations, we have assumed a viscosity
of $\nu =3\cdot 10^{-6}\,m^2/s$, which is at the low end of typical
measured blood viscosities ($\sim 3-5\cdot 10^{-6}\,m^2/s$, \cite{lan69}).
Thus, we have determined lower bounds of the instability thresholds in
Fig.~\ref{fig_stab} and 
the actual regions of stability may be somewhat larger. Note also that
stability will be enhanced when lower driving frequencies are used as
this leads to a larger effective damping of surface oscillations
(cf.\ \cite{bre95}).

%This may be the reason why preliminary experiments at Schering's
%with water failed to show higher sound harmonics.
%To detect a sound signal of the second harmonic
%we suggest to perform experiments with blood or a liquid with a correspondingly
%enlarged viscosity.
Stability analysis shows that very large driving amplitudes (say 10\,atm)
will not help provide large response signals from the suspended bubbles.
Instead, it may be useful to limit the amplitudes in the bubble regions
to $\aleq2.5$\,atm.

The predictions for the region of optimal $I_2/I_1$ intensity ratio remain
virtually unaltered, as there is very little overlap of this region with
the instability tongues.
Indeed, according to the regime of large $I_2/I_1$ identified above,
shape
instability in this regime is to be expected only in a tiny
area of parameter space at $R_0\approx 1.2\,\mu m$ and $P_a\approx 3\,$atm
(see Fig.~\ref{fig_stab}\,b).

Besides spherical instability, diffusive instability and chemical
instability also are matters of concern, as
pointed out in detail for 26\,kHz forced bubbles
in refs.\ \cite{hil96,loh96}. Both types of instabilities
will only be important on long timescales
(milliseconds or longer) where our approximation
of a stable standing wave is not appropriate anyhow. This is why
we postpone the discussion of these instabilities to future work.

\section{Summary and conclusions}\label{conclsec}

Bubble suspensions as contrast enhancers in ultrasound diagnostics are now state
of the art. Improving the image quality by detecting the second harmonic of
the driving frequency in the emitted sound spectrum is likely to lead to
their further acceptance, as the advantages compared to conventional methods
become even more pronounced. We have identified regions in parameter space,
i.e., values for the driving pressure amplitude $P_a$ and the ambient bubble
radius $R_0$, where relatively high sound intensities at the second harmonic
frequency are to be expected. These regions are intimately connected to
the resonance structure of the bubble oscillator and to the collapse dynamics
of the bubble.
The best suited parameter regime 
to achieve a high absolute second harmonic intensity $I_2$ is located
around the main bubble resonance radius, i.e., 
$R_0=1.0 - 1.5\,\mu m$ if a fixed driving frequency
of 3\,MHz is used. Requiring bubble stability towards non-spherical perturbations
limits useful driving pressures to a maximum of about $2.5$\,atm, if the bubbles
oscillate in blood (cf.\ Fig.~\ref{fig_stab}b). 
Here, we employed the previously specified values for
 $\rho_l, c_l, \nu, \sigma$ to model the average properties of blood. 
Bubbles in this parameter range are stable 
in fluids with (at least) three times higher viscosity than water.

The intensity ratio $I_2/I_1$ is important for diagnostic purposes (switching
between ``background'' and contrast agent images). It is optimized for
bubbles around $R_0\approx 0.8\,\mu m$ and $P_a\approx1-3$\,atm. Note that, again,
there is an upper limit to the strength of optimal driving. 

We therefore suggest not to use
very high ($\geq 2.5$\,atm) pressure amplitudes;
%(as is now common practice with peak pressures up to 10\,atm);
a gentler driving may lead to a better image quality.
Also, the bubble radii should be somewhat
smaller than those dominant in bubble suspensions used today (e.g.\ SH U 508 A
with a radius distribution peak at $\approx 1.4\,\mu m$ \cite{nan93}), if
maximum efficiency at 3\,MHz driving frequency is to be achieved. A narrower
distribution around the peak (i.e., more uniform radii) would, of course,
further amplify  the sound signal.

\vspace{0.5cm}

\noindent
{\bf Acknowledgments:}
The authors thank T. Matula for valuable comments on the manuscript.
This work has been supported by the DFG through its
SFB185.

%\vspace{-0.2cm}

%\newpage

\end{document}